\documentclass[10pt,a4paper,twoside]{article}
\usepackage{epsfig}
\usepackage{baltlat5}
\usepackage{wrapfig}

\begin{document}
\ \
\vspace{0.5mm}

\setcounter{page}{1}
\vspace{5mm}

\titlehead{Baltic Astronomy, vol.\ts 14, XXX--XXX, 2005.}

\titleb{LINE-PROFILE VARIATIONS IN PULSATING SUBDWARF-B\\ STARS AS A PULSATION MODE DIAGNOSTIC}

\begin{authorl}
\authorb{C.~Schoenaers}{} and
\authorb{A.~E.~Lynas-Gray}{}
\end{authorl}

\moveright-3mm
\vbox{
\begin{addressl}
\addressb{}{Department of Physics, University of Oxford, Denys Wilkinson Building, Keble Road, Oxford OX1 3RH, United Kingdom}
\end{addressl}
}
\submitb{Received 2005 July 31}

\begin{summary}
In previous attempts to perform seismic modelling of pulsating subdwarf-B stars, various mode identification techniques are used with uncertain results.

We investigated a method so far neglected in sdB stars, but very successful for Main Sequence pulsators, that is, mode identification from the line-profile variations caused by stellar pulsation.

We report the calculation of time-resolved synthetic spectra for sdB stars pulsating with various combinations of pulsation modes; these calculations were carried out over appropriate ranges of effective temperature, surface gravity and helium abundances. Preliminary tests using these synthetic line-profile variations demonstrated their potential for mode identification by comparison with observation.
\end{summary}


\resthead{LPV in pulsating sdB stars as a pulsation mode diagnostic}{C.~Schoenaers \&\ A.~E.~Lynas-Gray}

\sectionb{1}{INTRODUCTION}
Subdwarf B (sdB) stars are believed to be low-mass $(\sim 0.5 M_{\odot})$ core helium 
burning objects belonging to the Extreme Horizontal Branch (EHB) (Heber~1986).  With thin and mostly 
inert hydrogen-rich residual envelopes, sdB stars remain hot 
$(20000 \le T_{\rm eff} \le 40000)$ and compact $(5 \le {\log g} \le 7)$ throughout 
their EHB lifetime (Saffer {\it et al.} 1994); they eventually evolve towards the white dwarf cooling sequence without 
experiencing the Asymptotic Giant Branch and Planetary Nebula phases of stellar evolution. 
While binary population synthesis calculations by Han {\it et al.} (2002, 2003) 
successfully demonstrate the formation of sdB stars through several possible channels, 
resulting models require further comparison with observation.  As sdB stars are 
believed to be responsible for the ultraviolet-upturn seen in the energy distributions 
of elliptical galaxies and spiral galaxy bulges (Yi {\it et al.} 1997), understanding sdB 
star formation may provide an important diagnostic for studying galaxy evolution and 
formation. 

The discovery (Kilkenny {\it et al.} 1997, Green {\it et al.} 2003) that some sdB stars 
are nonradial pulsators means asteroseismology could be used to discern their internal 
structure and so constrain evolution models. But before any seismic modelling can be reliably attempted one must identify pulsation modes which are excited. 

\begin{wrapfigure}[15]{r}[0pt]{80mm}
\vskip-1mm
\centerline{\psfig{figure=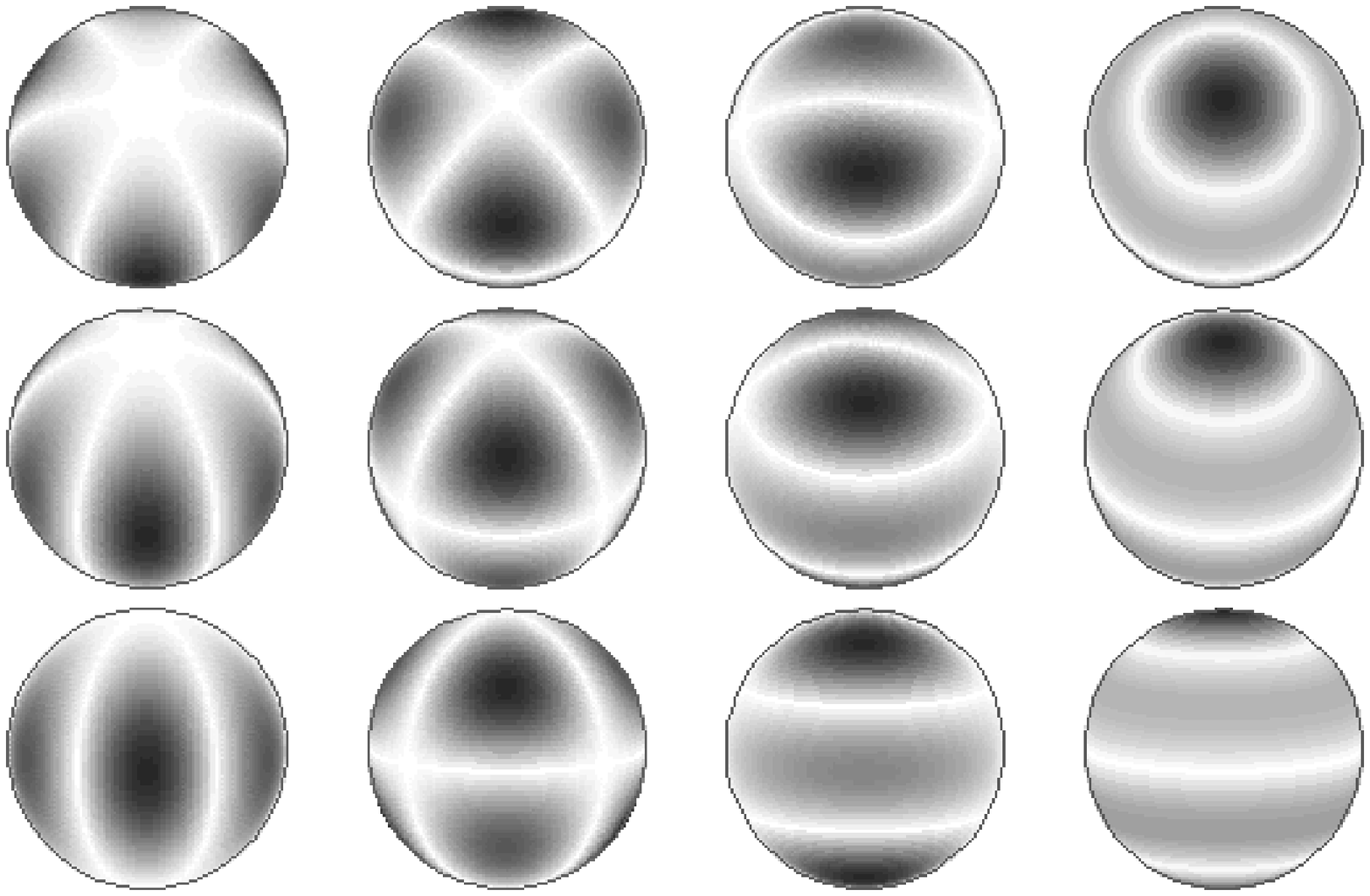,width=70mm,clip=}}
\vskip1mm
\captionb{1}{$l$\,=\,3 pulsation mode with $m$ from 3 (left) to 0 (right) and $i$ from $30\degr$ (top) to $90\degr$ (bottom) (GONG/NSO/AURA/NSF).}
\end{wrapfigure}

Assuming radial and non-radial pulsations can be modelled by spherical harmonics $Y_{nlm}$, mode identification tries to assign values to the spherical wavenumbers $n$, $l$ and $m$: $n$ is related to the number of nodes of the radial displacement, $l$ is the number of nodal lines on the stellar surface and $m$ is the number of such lines passing through the rotation axis of the star, as shown in Fig.~1.

\sectionb{2}{THE PROBLEM OF MODE IDENTIFICATION IN SDB STARS}
Mode identification in subdwarf-B stars is far from being trivial. Several methods which prove efficient for mode identification in pulsating Main Sequence stars have limitations in the case of sdB stars. Nevertheless, these mode identification techniques, however uncertain, are used in meritorious attempts at seismic modelling of the most promising pulsating sdB stars.

\subsectionb{2.1}{Period fitting}
The method entails fitting an observed period spectrum of a pulsating star to a theoretical period spectrum and is straightforward in principle. However, when considering pulsating sdB stars there is a major difficulty to overcome (Brassard {\it et al.} 2001): so far, most sdB stars (except for PG\,1605+072 with more than 50 periods (Koen {\it et al.} 1998), and KPD\,1930+2752 with at least 44 periods (Bill\`{e}res {\it et al.} 2000)) have a rather sparse period spectrum, possibly because lower amplitude frequencies are not yet observed. This makes it very difficult to match objectively observed periods with periods computed from a model, as the latter are much more numerous.

Despite the intrinsic difficulty of period fitting in sdB stars, Brassard {\it et al.} (2001) use this technique to model the pulsating sdB star PG\,0014+067 and tentatively identify 23 distinct pulsation modes. They search a four-parameter space ($T_\mathrm{eff}$, $\log g$, the mass of the star $M_\ast$ and the mass of its H-rich envelope $M_\mathrm{env}$) for a model whose theoretical period spectrum could account for the observed periods in PG\,0014+067, using a ``merit" function that is minimum for the best fit(s). In addition to the four parameters, they also deduce the rotation velocity of the star and other properties (see their Table 6). They claim a good agreement with previous spectroscopic determination of $T_\mathrm{eff}$ and $\log g$, but we note that in order to discriminate between equally good fits (several minima of their merit function) they use spectroscopic constraints. More recently, the same method is applied by Charpinet {\it et al.} (2005) to PG\,1219+534.

\subsectionb{2.2}{Amplitude ratios}
Yet another method in use for mode identification in pulsating Main Sequence stars relies on multicolor photometry. This method, known as the amplitude ratio method (Heynderickx {\it et al.} 1994), relies on the fact that for non-radially pulsating stars the photometric pulsation amplitude depends on wavelength and on the spherical wavenumber $l$ of the pulsation mode. Ramachandran {\it et al.} (2004) investigate this method for non-radially pulsating sdB stars and draw the following conclusions:
\begin{itemize}
\item for EC\,14026 stars, only the modes of high $l$ (3 and 4) should be easy to disentangle, provided their amplitudes are large enough to observe. On the other hand, low degree modes are very difficult to distinguish, and this problem becomes worse as the period shortens;
\item for ``Betsy" stars, modes of spherical degree $l=1$ may be difficult to distinguish from modes with $l>1$ in the case of longer periods.
\end{itemize}

Using {\sc ultracam} multicolor photometry, Jeffery {\it et al.} (2004, 2005) use the amplitude ratio method to identify pulsation modes of three sdB stars, HS\,0039+4302, KPD\,2109+4401 and PG\,0014+067. Their results are provisional, as only statistical errors are taken into account.

\sectionb{3}{LINE-PROFILE VARIATIONS IN PULSATING SDB STARS}
We saw in the previous section that current mode identification methods in sdB stars need improvement. However none of these techniques use spectroscopy to perform mode identification, although it has proved useful when modelling pulsating Main Sequence stars (e.g. Aerts {\it et al.} 1992, Briquet \& Aerts 2003). Our claim is that spectroscopy, and the pioneering detailed computation of line-profile variations (lpv) we perform, could provide a more reliable mode identification method. Indeed lpv enable a complete reconstruction of the non-radial oscillations of a star, provided that a time-series of spectra are observed with a high enough S/N ratio, temporal, and wavelength resolution.

\subsectionb{3.1}{Computation of synthetic line-profile variations}
The method used for our computation of theoretical line-profile variations is quite straightforward. We combined three computer programs, BRUCE (Townsend 1997), SYNSPEC (Hubeny \& Lanz 2000) and KYLIE (Townsend 1997), so as to compute a series of time-resolved spectra exhibiting line-profile variations. 

BRUCE first divides the stellar surface into a large number of very small surface elements that are determined by the usual polarand azimuthal coordinate angles $\theta$ and $\varphi$ and a step-size $\Delta\theta$ and $\Delta\varphi$. BRUCE then perturbs the stellar surface, assuming the pulsation is linear and adiabatic, with selected combinations of pulsation modes at selected time steps, giving for each surface element its temperature, surface gravity, orientation, projected radial velocity and projected area.

It is important to realize that, because of the pulsation, different points on the stellar surface not only have different radial velocity, but also different temperature, $\log g$ and orientation, meaning that the contributions of the different surface elements to the line-profile have different amplitudes and energy distributions. The line-profile variations do not only come from the Doppler shift in wavelength, but also from the complex temperature and $\log g$ behavior on the stellar surface.


The emergent monochromatic flux in wavelength interval $\Delta \lambda$, at some wavelength $\lambda$ and time $t$ is then given by
\begin{equation}\label{lp}
F_\lambda (t)=\sum_{i,j}I_\lambda(t)_{ij}\,\mu(t)_{ij}\,\sin\theta_i\,\Delta\theta_i\,\Delta\varphi_j
\end{equation}
where $I_\lambda(t)_{ij}$ is the specific intensity emergent from the tile at $\theta_i$ and $\varphi_j$ in wavelength interval $\Delta\lambda$, at wavelength $\lambda$ and time $t$, in the direction of the observer, as computed by our modified version of SYNSPEC. The quantity
\begin{displaymath}
\mu(t)_{ij}\equiv\vec{n}(t)_{ij}\cdot\vec{s}
\end{displaymath}
accounts for the time-dependent orientation of each tile, $\vec{n}(t)_{ij}$ being the normal to the tile at some time $t$, and $\vec{s}$ being the direction of propagation of the radiation.

This is then repeated for the desired number of time-steps using KYLIE, therefore producing series of time-resolved observer-directed spectra.

Examples of line-profile variations computed with our codes are given for different modes in Figs.~2 and 3.

\begin{figure}[t!]
\centerline{\psfig{figure=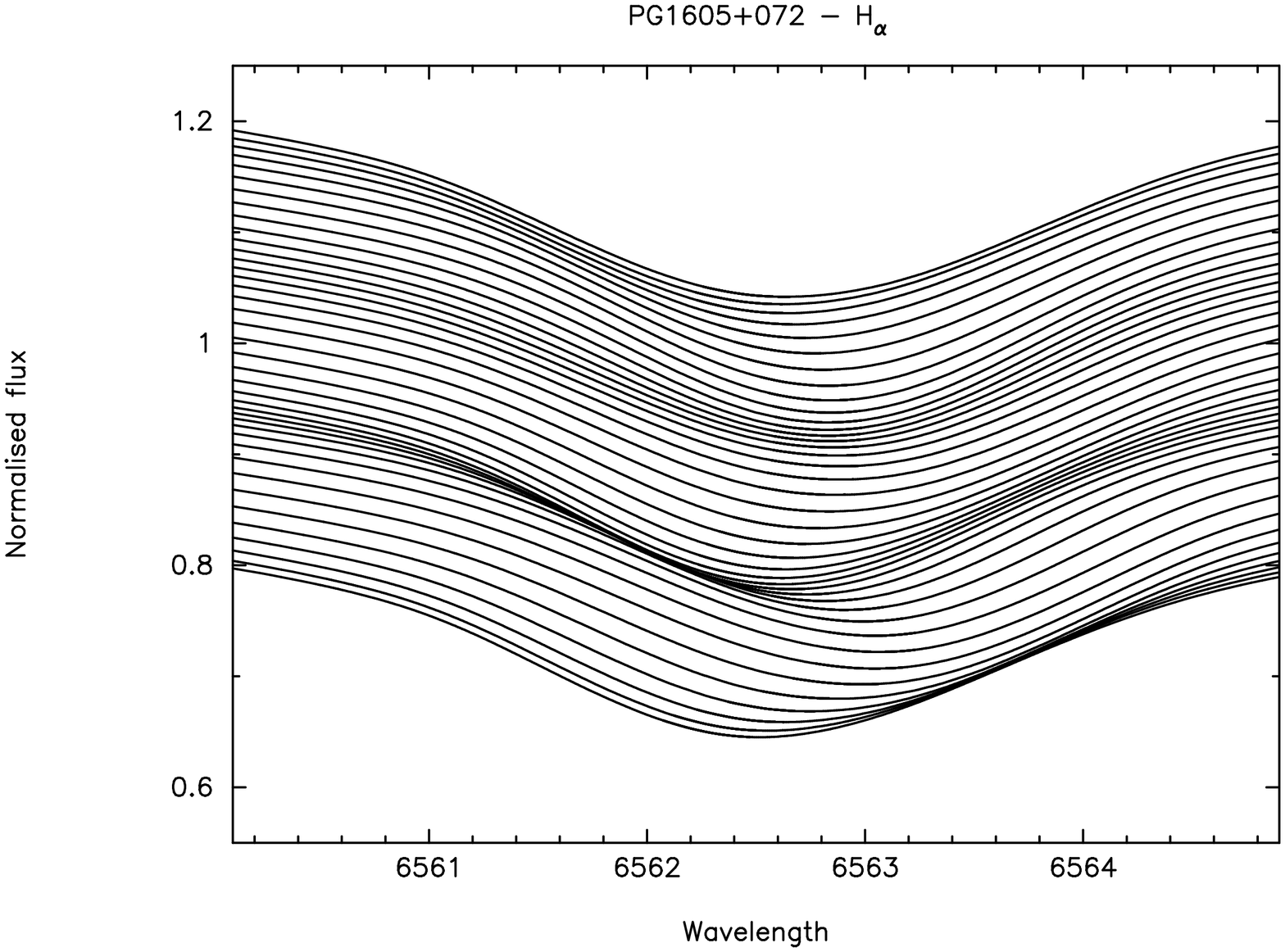,width=8cm,angle=270,clip}}
\captionb{2}{Trial computation of H$_\alpha$ line-profile variations for the short-period pulsator PG\,1605+072. Only the six highest-amplitude pulsation modes (out of more than 50) were taken into account in this computation in order to reduce computing time. The line-profiles have been shifted along the ordinate axis (normalized flux) for clarity.}
\end{figure}

\begin{figure}[!t]
\centerline{\psfig{figure=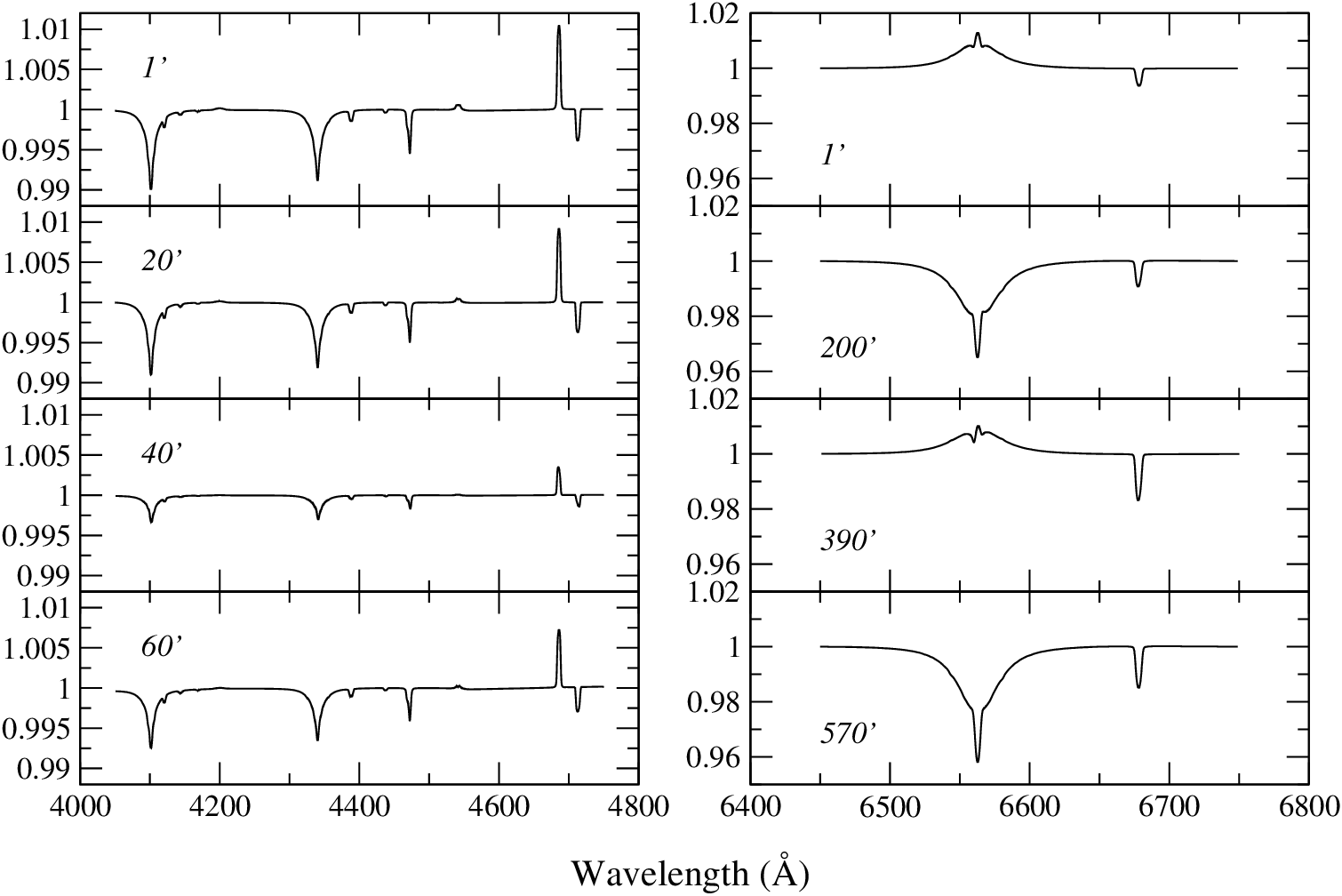,width=11cm,clip}}
\captionb{3}{A time series of 1\AA\ resolution synthetic spectrum ratios for 
PB\,8783 (short-period pulsator) and EC\,21324-1346 (long-period pulsator) in left and right frames respectively. Second and 
subsequent synthetic spectra in each time series are divided by the first, 
the elapsed time being given in minutes. Time-series calculations were based on assumed 
pulsation modes $(l=3,m=0;l=2,m=0;l=1,m=0;l=2,m=1;l=3,m=0)$ and velocity amplitudes from Jeffery \& Pollacco (2000) for PB\,8783 and on modes $(l=1,m=1;l=2,m=1;l=3,m=0)$ and amplitudes to match photometric variations of EC\,21324$-$1346.} 
\end{figure}

\subsectionb{3.2}{Observation of line-profile variations in pulsating sdB stars}
Observations of line-profile variations in sdB stars have been made by Telting \& \O stensen (2004) who observed PG\,1325+101 with the 2.5-m NOT telescope. Their observations showed strong evidence for a dominant radial pulsation mode, but most 
importantly drew attention to the potential of spectroscopic mode identification. Similar observations were made of the bright sdB star Balloon 090100001 (Telting \& \O stensen 2005).

\sectionb{4}{THE MOMENT METHOD AND ITS APPLICATION TO SDB STARS}
\subsectionb{4.1}{The moment method}
Mode identification from line-profile variations could in principle be carried out 
by a direct comparison of observed and synthetic spectra but we propose to adapt to pulsating sdB stars the so-called \emph{moment method}. It was first introduced by Balona (1986a,b, 1987), and later refined and applied to Main Sequence pulsators by Aerts {\it et al.} (1992), Aerts (1996) and Briquet \& Aerts (2003). 

In this method, an observed line-profile is replaced by its first few moments, and their time dependence studied to identify the pulsation modes. The $n$th moment of a line-profile $p(v)$ (where $v$ is the line of sight velocity component corresponding to the displacement from the laboratory wavelength) is be defined as
\begin{equation}
<v^n>\equiv \frac{\int^{+\infty}_{-\infty}v^np(v)\,dv}{\int^{+\infty}_{-\infty}p(v)\,dv}.
\end{equation}
All the information contained in the line-profile can be reconstructed from the entire series of moments of $n$th order, but Briquet \& Aerts (2003) showed that most of the information is contained in the first three moments:
\begin{itemize}
\item the first moment $<v>$ is the centroid of the line-profile in a reference frame with origin at the stellar center;
\item the second moment $<v^2>$ is related to the variance of the line-profile;
\item the third moment $<v^3>$ relates to the skewness of the profile (see also Telting 2003).
\end{itemize} 

\begin{wrapfigure}[25]{r}[0pt]{60mm}
\vskip-3mm
\centerline{\psfig{figure=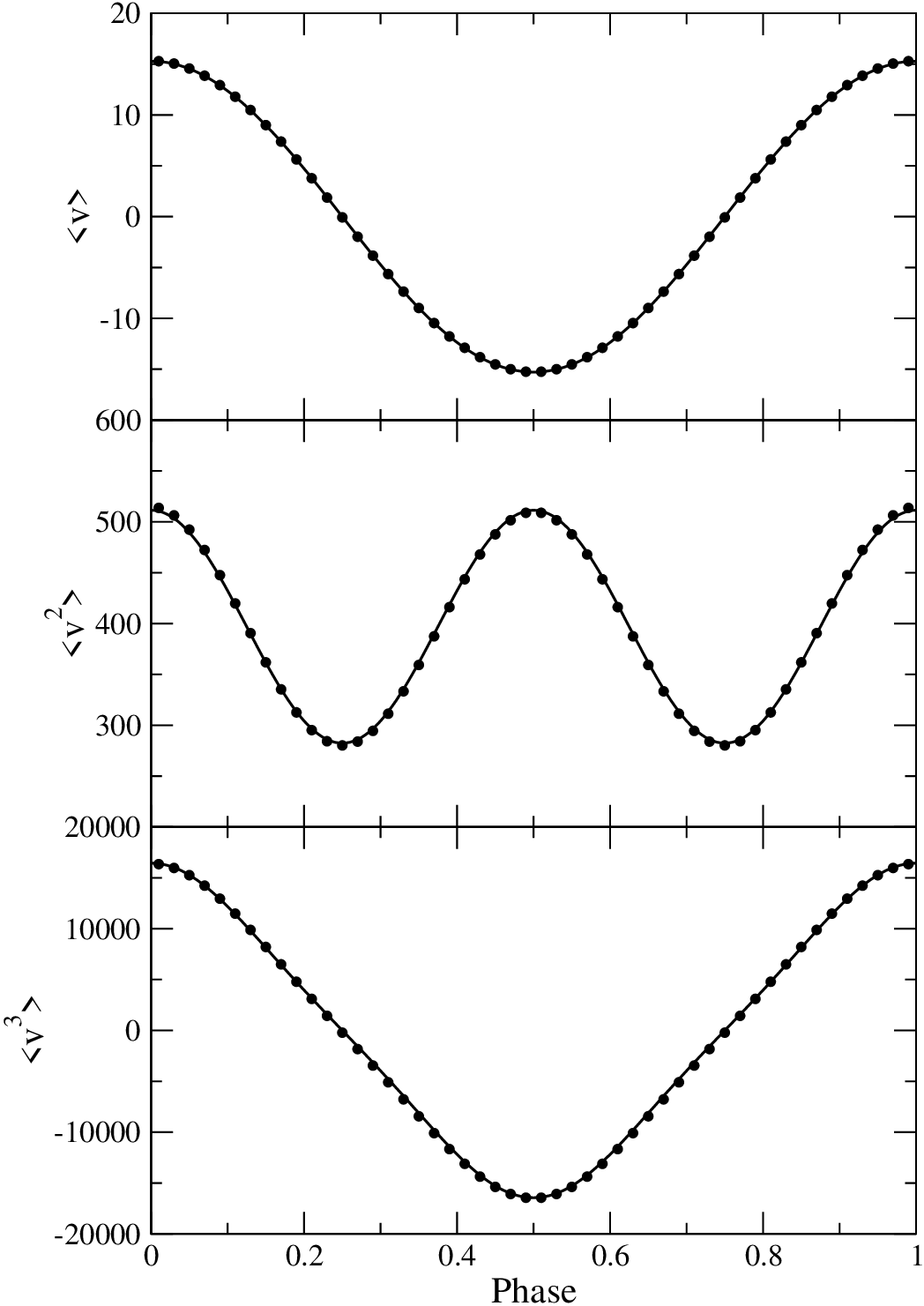,width=59mm,clip}}
\vskip-1mm
\captionb{4}{First, second and third moments for the synthetic line profile of a radial mode computed using our codes.}
\end{wrapfigure}

Once the moments of the observed line-profile have been computed, their variations are compared with the time-dependence of theoretical moments (Aerts 1992). These depend on the wavenumbers $l$ and $m$ of the pulsation mode, but also on its amplitude $v_p$, on the inclination $i$ of the star's axis of rotation to the observer, on its rotation velocity $v_\Omega$ and on the width $\sigma$ of the intrinsic broadening profile, allowing a complete reconstruction of non-radial pulsation. One then selects the most likely set of parameters $(l,m,v_p,i,v_\Omega,\sigma)$ by minimizing a so-called discriminant (Briquet \& Aerts 2003), which is a function that takes into account the quality of the fit between the theoretical and observed moments.

In Fig.~4, we show the three normalized moments for a synthetic radial mode we computed.  
Even without computing the discriminant, the phase-dependence of the observed moments can provide some useful information. For instance 
\begin{itemize}
\item the peak-to-peak amplitude of the first moment gives an idea about the overall velocity range due to the oscillation;
\item if $<v^2>$ can be described by a single sine function with twice the frequency of the first moment it means that $m=0$, while if the mode is sectoral ($|m|=l$) $<v^2>$ behaves sinusoidally with the same frequency as the frequency of the first moment.
\end{itemize}

\subsectionb{4.2}{Application to pulsating sdB stars}
The major difficulty to overcome in order to apply the moment method to pulsating sdB stars comes from its current formulation (Briquet \& Aerts 2003) in which the moment method neglects the variation of the specific intensity (i.e. $\delta I_\lambda(\theta_i,\varphi_j)=0$) due to changes in $T_\mathrm{eff}$ and $\log g$ during the pulsation. This can be a good approximation when dealing with spectral lines that are not sensitive to temperature variations (such as silicon lines in $\beta$ Cephei stars (Dupret {\it et al.} 2002) and (De Ridder {\it et al.} 2002)) but in the case of sdB stars, temperature effects should not be neglected. 

We tested the moment method in its current formulation on lpv of H$_\alpha$ for various pulsation modes, but the identification was not always satisfactory. However it was extremely difficult to choose a suitable wavelength range over which the moments were to be computed. We are investigating a systematic way to make that choice, as well as whether the use of the narrower and much weaker helium lines might improve the mode identification. Another possibility is to take temperature effects into account using of higher-order moments, as Balona (1987) suggests.

\sectionb{5}{FUTURE PROSPECTS}
Our development of spectroscopic methods for pulsation mode identification in subdwarf-B stars was stimulated by the acquisition of observational data. A start on this was made in August 2005 using the South African Astronomical Observatory 1.9-m telescope to obtain time-resolved spectroscopy of ``Betsy" stars (Schoenaers \& Lynas-Gray 2006).

Furthermore, Kawaler \& Hostler (2005) show internal rotation to be expected in sdB stars, the rate 
depending on distance from the stellar centre.  Mode splitting due internal rotation 
cannot be identified using the usual formula applicable in the case of surface rotation 
and may have led to incorrect mode identification in previous works. Special attention therefore needs to be given to the modelling of this effect in pulsating sdB stars, even though the moment method shouldn't be affected much by differential rotation, contrarily to period fitting.

However, mode identification in pulsating sdB stars is a real challenge with the methods discussed above. Other methods might prove worth investigating.

The so-called IPS method aims at comparing the observed amplitude and phase variations across the line-profile with the theoretical variations (Schrijvers {\it et al.} 1997 and Telting \& Schrijvers 1997a, b). It was recently used by De Cat {\it et al.} (2005) to identify pulsation modes in monoperiodic SPB (Slowly Pulsating B) stars, together with the method of photometric amplitude ratios and the moment method (with moments up to the sixth order).

Yet another mode identification method, Doppler Imaging, should be investigated. The method is based on the idea that line-profile variations of rapidly rotating stars are a sort of ``doppler image" of the stellar surface within the line-profile. This method has been challenged because of its apparent lack of physical foundation, but Kochukhov (2004) claims a reliable recovery of the surface pulsation velocity structures can be achieved for all types of pulsation geometries accompanied by significant line-profile variations. It remains to be seen whether the line-profile variations in sdB stars are large enough.

To summarize, this paper presents several options for meeting the challenge presented by sdB star pulsation mode identification by spectroscopic methods. All appear to be worthy of further investigation and progress made to date is briefly summarized.

\vskip5mm

ACKNOWLEDGMENTS. The authors are indebted to Conny Aerts for a stimulating discussion, to John Telting for his valuable comments as referee and the University of Oxford, Department of Physics, for travel grants.

\References

\refb
Aerts~C. 1996, A\&A 314, 115
\refb
Aerts~C., de Pauw~M., Waelkens~C. 1992, A\&A 266, 294 
\refb
Balona~L.~A. 1986a, MNRAS 219, 111
\refb
Balona~L.~A. 1986b, MNRAS 220, 647
\refb
Balona~L.~A. 1987, MNRAS 224, 41
\refb
Bill\`{e}res~M., Fontaine~G., Brassard~P.~et al.
   2000, ApJ 530, 441
\refb
Brassard~P., Fontaine~G., Bill\`{e}res~M.~et al.
   2001, ApJ 563, 1013
\refb
Briquet~M., Aerts~C. 2003, A\&A 398, 687
\refb
Charpinet~S., Fontaine~G., Brassard~P., Green~E., Chayer~P. 2005, A\&A 437, 575
\refb
De Cat~P., Briquet~M., Daszynska-Daszkiewicz~J.~et al.
   2005, A\&A 432, 1013
\refb
De Ridder~J., Dupret~M.-A., Neuforge~C., Aerts~C. 2002, A\&A 385, 572
\refb
Dupret~M.-A., De Ridder~J., Neuforge~C., Aerts~C. 2002, A\&A 385, 563
\refb
Green~E.~M., Fontaine~G., Reed~M.~D.~et al.
   2003, ApJ 583, L31
\refb
Han~Z., Podsiadlowski~Ph., Maxted~P.~F.~L., Marsh~T.~R. 2003, MNRAS 341, 669
\refb
Han~Z., Podsiadlowski~Ph., Maxted~P.~F.~L., Marsh~T.~R., Ivanova~N. 2002, MNRAS 336, 449
\refb
Heber~U. 1986, A\&A 155, 33
\refb
Heynderickx~D., Waelkens~C., Smeyers~P. 1994, A\&AS 105, 447
\refb
Hubeny~I., Lanz~T. 2000, \texttt{http://tlusty.gsfc.nasa.gov/Synspec43/synspec.html}
\refb
Jeffery~C.~S., Aerts~C., Dhillon~V.~S., Marsh~T.~R. 2005, in {\it Hot Subdwarf Stars and Related Objects}, BA JJJ, p. PPP
\refb
Jeffery~C.~S., Dhillon~V.~S., Marsh~T.~R., Ramachandran~B. 2004, MNRAS 352, 699
\refb
Jeffery~C.~S., Pollacco~D. 2000, MNRAS 318,974
\refb
Kawaler~S.~D., Hostler~S.~R. 2005, ApJ 621, 432
\refb
Kilkenny~D., Koen~C., O'Donoghue~D., Stobie~R.~S. 1997, MNRAS 285, 640
\refb
Kochukhov~O. 2004, A\&A 423, 613
\refb
Koen~C., O'Donogue~D., Kilkenny~D.~et al.
   1998, MNRAS 296, 317
\refb
Ramachandran~B., Jeffery~C.~S., Townsend~R.~H.~D. 2004, A\&A 428, 209
\refb
Saffer~R.~A., Bergeron~P., Koester~D., Liebert~J. 1994, ApJ 432, 351
\refb
Schoenaers~C., Lynas-Gray~A.~E. 2006, MNRAS, {\it in preparation}
\refb
Schrijvers~C., Telting~J.~H., Aerts~C., Ruymaekers~E., Henrichs~H.~F. 1997, A\&AS 121, 343
\refb
Telting~J.~H. 2003, Ap\&SS 284, 85
\refb
Telting~J.~H., \O stensen~R.~H. 2004, A\&A 419, 685
\refb
Telting~J.~H., \O stensen~R.~H. 2005, in {\it Hot Subdwarf Stars and Related Objects}, BA JJJ, p. PPP
\refb
Telting~J.~H., Schrijvers~C. 1997a, A\&A 317, 723
\refb
Telting~J.~H., Schrijvers~C. 1997b, A\&A 317, 742
\refb
Townsend~R.~H.~D. 1997, {\it Ph.D. Thesis}, University College London
\refb
Yi~S., Demarque~P., Oemler~A. 1997, ApJ 486, 201

\end{document}